\documentclass[a4paper]{article}
\pdfoutput=1

\usepackage{amsmath}
\usepackage{authblk}
\usepackage{graphicx}
\usepackage{ifpdf}
\usepackage[sort&compress,round]{natbib}
\usepackage{subfigure}
\usepackage{url}
\usepackage{xspace}

\ifpdf
	\usepackage{pdfsync}
\else
	\usepackage{draftcopy}
\fi

\usepackage{hyperref}

\newcommand{\wraprefprepost}[3]{\wrapprepost{#1}{#2}{\ref{#3}}}
\newcommand{\formatrefplain}{\ref}
\newcommand{\formatrefparens}{\wraprefprepost{(}{)}}

\newcommand{\wrapprepost}[3]{{#1}{#3}{#2}}

\newcommand{\tagwithlabel}[2]{#1~#2}

\newcommand{\makelabeledcrossrefmacro}[4]
	{\newcommand{#3}{#1{#4}{#2}}}
\newcommand{\makecrossrefmaker}[3]
	{\newcommand{#1}{\makelabeledcrossrefmacro{#2}{#3}}}

\newcommand{\eqnrefformat}{\formatrefparens}
\newcommand{\eqnlabelbinding}{\tagwithlabel}
\newcommand{\eqnlabel}{eq.}
\newcommand{\Eqnlabel}{Eq.}
\newcommand{\eqnslabel}{eqs.}
\newcommand{\Eqnslabel}{Eqs.}

\newcommand{\eqnnum}{\eqnrefformat}
\makecrossrefmaker{\newlabeledeqnref}{\eqnlabelbinding}{\eqnnum}
\makecrossrefmaker{\newwordpluseqnref}{\tagwithlabel}{\eqnnum}

\newlabeledeqnref{\eqn}{\eqnlabel}
\newlabeledeqnref{\Eqn}{\Eqnlabel}
\newlabeledeqnref{\eqns}{\eqnslabel}
\newlabeledeqnref{\Eqns}{\Eqnslabel}

\newwordpluseqnref{\andeqn}{and}
\newwordpluseqnref{\througheqn}{through}

\newcommand{\figrefformat}{\formatrefplain}
\newcommand{\figlabelbinding}{\tagwithlabel}
\newcommand{\figlabel}{fig.}
\newcommand{\Figlabel}{Fig.}
\newcommand{\figslabel}{figs.}
\newcommand{\Figslabel}{Figs.}

\newcommand{\fignum}{\figrefformat}
\makecrossrefmaker{\newlabeledfigref}{\figlabelbinding}{\fignum}
\makecrossrefmaker{\newwordplusfigref}{\tagwithlabel}{\fignum}

\newlabeledfigref{\fig}{\figlabel}
\newlabeledfigref{\Fig}{\Figlabel}
\newlabeledfigref{\figs}{\figslabel}
\newlabeledfigref{\Figs}{\Figslabel}

\newwordplusfigref{\andfig}{and}
\newwordplusfigref{\throughfig}{through}

\newcommand{\sxnrefformat}{\formatrefplain}
\newcommand{\sxnlabelbinding}{\tagwithlabel}
\newcommand{\sxnlabel}{section}
\newcommand{\Sxnlabel}{Section}
\newcommand{\sxnslabel}{sections}
\newcommand{\Sxnslabel}{Sections}

\newcommand{\sxnnum}{\sxnrefformat}
\makecrossrefmaker{\newlabeledsxnref}{\sxnlabelbinding}{\sxnnum}
\makecrossrefmaker{\newwordplussxnref}{\tagwithlabel}{\sxnnum}

\newlabeledsxnref{\sxn}{\sxnlabel}
\newlabeledsxnref{\Sxn}{\Sxnlabel}
\newlabeledsxnref{\sxns}{\sxnslabel}
\newlabeledsxnref{\Sxns}{\Sxnslabel}
	
\newwordplussxnref{\andsxn}{and}
\newwordplussxnref{\throughsxn}{through}

\newcommand{\tblrefformat}{\formatrefplain}
\newcommand{\tbllabelbinding}{\tagwithlabel}
\newcommand{\tbllabel}{table}
\newcommand{\Tbllabel}{Table}
\newcommand{\tblslabel}{tables}
\newcommand{\Tblslabel}{Tables}

\newcommand{\tblnum}{\tblrefformat}
\makecrossrefmaker{\newlabeledtblref}{\tbllabelbinding}{\tblnum}
\makecrossrefmaker{\newwordplustblref}{\tagwithlabel}{\tblnum}

\newlabeledtblref{\tbl}{\tbllabel}
\newlabeledtblref{\Tbl}{\Tbllabel}
\newlabeledtblref{\tbls}{\tblslabel}
\newlabeledtblref{\Tbls}{\Tblslabel}
	
\newwordplustblref{\andtbl}{and}
\newwordplustblref{\throughtbl}{through}

\newcommand{\etc}{etc.} 
\newcommand{\eg}{e.g.}
\newcommand{\brim}{BRIM\xspace}
\newcommand{\fm}{FM\xspace}
\newcommand{\RD}{R\&{}D\xspace}
\newcommand{\CCM}{Centro de Ci\^encias Matem\'aticas\xspace}

\newcommand{\newterm}{}
\newcommand{\subjectindex}[1]{\textsl{#1}}

\newcommand{\acknowledgments}
	{\section*{Acknowledgments}\label{sec:acknowledgments}} 

\newcommand{\largeparens}[1]{\left ( #1 \right )}
\newcommand{\largesquare}[1]{\left [ #1 \right ]}

\newcommand{\largestraight}[1]{\left | #1 \right |}

\newcommand{\abs}{\largestraight}

\newcommand{\mathpuncspace}{\quad}
\newcommand{\mathcomma}{\mathpuncspace,}
\newcommand{\mathperiod}{\mathpuncspace.}

\newcommand{\mat}[1]{\boldsymbol{#1}}

\newcommand{\transpose}[1]{#1^\mathrm{T}}

\newcommand{\zeromat}{\mat{O}}

\newcommand{\vertexdegree}[1]{\ensuremath{d_{#1}}}

\newcommand{\adjmat}{\mat{A}}
\newcommand{\adjelem}[2]{A_{#1#2}}
\newcommand{\adjsubmat}{\mat{M}}

\newcommand{\probelem}[2]{P_{#1#2}}

\newcommand{\numedges}{\ensuremath{M}}
\newcommand{\modularity}{\ensuremath{Q}}

\newcommand{\freq}[1]{f\largeparens{#1}}
\newcommand{\topfreq}[2]{\topfreqnoarg{#1}\largeparens{#2}}
\newcommand{\topfreqnoarg}[1]{f_{#1}}

\newcommand{\matrixbrackets}{\largesquare}

\newcommand{\bipartsubstructure}[1]{\matrixbrackets{
	\begin{array}{cc}
		\zeromat & #1 \\
		\transpose{#1} & \zeromat
	\end{array}}}

\newcommand{\topicmetric}[1]{d_{#1}}

\graphicspath{{pdffigs/}{epsfigs/}{mpsfigs/}}

\title{Community Structure and Topical Differentiation in European RTD Collaborations}

\author[1]{Michael J. Barber}
\author[2]{Margarida Faria}
\author[2]{Ludwig Streit}
\author[2]{Oleg Strogan}

\affil[1]{Austrian Research Centers GmbH---ARC, Bereich systems research, Vienna, Austria}
\affil[2]{\CCM, Universidade da Madeira, Funchal, Portugal}

\date{October 10, 2008}

\begin{document}

\maketitle

\begin{abstract}
We investigate research and development collaborations
under the EU Framework Programs (FPs) for Research and Technological
Development. The collaborations in the FPs give rise to bipartite
networks, with edges existing between projects and the organizations
taking part in them. A version of the modularity
measure, adapted to bipartite networks, is presented. Communities are
found so as to maximize the bipartite modularity. Projects in the resulting 
communities are shown to be topically differentiated.
\end{abstract}


\section{Introduction}\label{sec:intro}

The EU Framework Programs (FPs) for Research and Technological
Development were implemented to follow two main strategic objectives:
First, strengthening the scientific and technological bases of
European industry to foster international competitiveness and,
second, the promotion of research activities in support of other
EU policies. In spite of their different scopes, the fundamental
rationale of the FPs has remained unchanged. All FPs share a few
common structural key elements. First, only projects of limited
duration that mobilize private and public funds at the national
level are funded. Second, the focus of funding is on multinational
and multi-actor collaborations that add value by operating at the
European level. Third, project proposals are to be submitted by
self-organized consortia and the selection for funding is based on
specific scientific excellence and socio-economic relevance criteria
\citep{RoeBar:2006b}. By considering the constituents of these
consortia, we                     can represent and analyze the FPs
as networks of projects and organizations. The resulting networks
are of substantial size, including over 50 thousand projects and
over 30 thousand organizations.

We have general interest in studying a real-world network of large
size and high complexity from a methodological point of view.
Furthermore,  socio-economic research emphasizes the central
importance of collaborative activities in \RD for economic
competitiveness \citep[see, for instance, ][among many others]{FagMowNel:2005}. 
Mainly for reasons of data availability,
attempts to evaluate quantitatively the structure and function of
the large social networks generated in the EU FPs have begun only
in the last few years, using social network analysis and complex
networks methodologies \cite{AlmOliLopSanMen:2007,BarFarStrStr:2008,BarKruKruRoe:2006,BreCus:2004,RoeBar:2008}. Studies to date point to the
presence of a dense and hierarchical network. A highly connected
core of frequent participants, taking leading roles within consortia,
is linked to a large number of peripheral actors, forming a giant
component that exhibits the characteristics of a small world.  

Networks have attracted a burst of attention in the last decade \citep[for useful reviews, see][]{ChrAlb:2007,DorMen:2004,New:2003,AlbBar:2002},
with applications to natural, social, and technological networks.
Of great current interest is the identication of community structure
within networks. Stated informally, a community is a portion of the
network whose members are more tightly linked to one another than
to other members of the network. A variety of approaches
\citep{AngBocMarPelStr:2007,GolKog:2006,Has:2006,NewLei:2007,ReiBor:2006,PalDerFarVic:2005,NewGir:2004,ClaNewMoo:2004,GirNew:2002}
have been taken to explore this concept; 
\citet{DanDiaDucAre:2005} and \citet{New:2004b} provide useful reviews. 
Detecting the community structure
allows quantitative investigation of relevant heterogeneous
substructures formed in the network.

We investigate networks of
research and development collaborations under the FPs. The
collaborations in the FPs give rise to bipartite networks, with
edges existing between projects and the organizations taking part
in them. With this construction, participating organizations are
linked only through joint projects. The resulting networks are quite
large compared to typical social networks, containing tens of
thousands of vertices and edges. At this scale, visualization of
the networks is quite difficult, so we instead take an algorithmic
approach to community identification. A version of the modularity
measure \citep{NewGir:2004}, adapted to bipartite networks \citep{Bar:2007}, 
is used to assess the quality of a division of the vertices into 
communities. Communities are
found by maximizing the bipartite modularity. We consider topical differentiation
of the communities found.


The rest of the paper is structured as follows. In \sxn{sec:dataprep}, we discuss the data used on
the FPs, continuing with definition of networks from the data in \sxn{sec:defnetworks}. We present
in \sxn{sec:commstruct} a summary of methods for identifying network communities, and apply the
methods to the FP networks in \sxn{sec:fpcommunities}. Finally, we discuss the consequences of
our findings in \sxn{sec:discussion} .

\section{Data Preparation} \label{sec:dataprep}

We draw on the latest version of the sysres EUPRO database. This
database includes all information publicly available through the
CORDIS projects database\footnote{\url{http://cordis.europa.eu}}
and is maintained by ARC systems research (ARC sys). The sysres EUPRO database presently 
comprises data on funded research
projects of the EU FPs (complete for FP1--FP5, and about 70\% complete for
FP6) and all participating organizations. It contains systematic
information on project objectives and achievements, project costs,
project funding and contract type, as well as information on the participating
organizations including the full name, the full address and the
type of the organization. 

For purposes
of network analyses, the main challenge is the inconsistency of the
raw data. Apart from incoherent spelling in up to four languages
per country, organizations are labelled inhomogeneously. Entries
may range from large corporate groupings, such as EADS, Siemens and
Philips, or large public research organizations, like CNR, CNRS and
CSIC, to individual departments and labs.

Due to these shortcomings, the raw data is of limited use for meaningful
network analyses. Further, any fully automated standardization procedure is 
infeasible. Instead, a labor-intensive, manual data-cleaning process is used in building 
the database.  \Citet{RoeBar:2008} describe the data-cleaning process in detail; here, 
we restrict discussion to the steps of the process relevant to the present work. These are:
\begin{enumerate}
	\item Identification of unique organization name. Organizational
	boundaries are defined by legal control. Entries are assigned
	to appropriate organizations using the more recently available
	organization name. Most records are easily identified, but,
	especially for firms, organization names may have  changed
	frequently due to mergers, acquisitions, and divestitures.
	
	\item Creation of  subentities. This is the key step for
	mitigating the bias that arises from the different scales
	at which participants appear in the data set. Ideally, we use 
	the actual group or organizational unit that
	participates in each project, but this information is only
	available for a subset of records, particularly in the case
	of firms. Instead, subentities that operate in fairly coherent activity areas
	are pragmatically defined. 
	Wherever possible, subentities are identified at the second
	lowest hierarchical tier, with each subentity
	comprising one further hierarchical sub-layer. Thus,
	universities are broken down into faculties/schools,
	consisting of departments; research organizations are broken
	down into institutes, activity areas, \etc, consisting of
	departments, groups or laboratories; and conglomerate firms
	are broken down into divisions, subsidiaries, \etc{} Subentities
	can frequently be identified from the contact information
	even in the absence of information on the actual participating
	organizational unit. 
	Note that subentities may still vary considerably in scale.
	
	\item Regionalization. The data set has been regionalized
	according to the European Nomenclature of Territorial Units
	for Statistics (NUTS) classification system\footnote{NUTS is a 
	hierarchical system of regions used by the
	statistical office of the European Community for the production of
	regional statistics. At the top of the hierarchy are NUTS-0 regions
	(countries) below which are NUTS-1 regions and then NUTS-2 regions, \etc}, where possible 
	to the NUTS3 level. Mostly, this
	has been done via information on postal codes.

\end{enumerate}
Due to resource limitations, only 
organizations appearing more than thirty times in the standardization table for FP1--FP5 have 
thus far been processed. This could bias the results; however, the networks have a structure  
such that the size of the bias is quite low \citep{RoeBar:2008}. 

\section{Network Definition}  \label{sec:defnetworks}

Using the sysres EUPRO database, for each FP we construct a network
containing the collaborative projects and all organizational
subentities\footnote{We work exclusively at the 
subentity level, and will interchangably refer to organizations and subentities.} 
that are participants in those projects.    
An organization
is linked to a project if and only if the organization is a member
of the project.  Since an edge never exists between two organizations
or two projects, the network is bipartite.  The network edges are
unweighted; in principle, the edges could be assigned weights to
reflect the strength of the participation, but the  data needed  to
assign the network weights is not available.

Previous investigations of the FPs often have made use of 
one-mode projection networks 
\citep{AlmOliLopSanMen:2007,BarKruKruRoe:2006,BreCus:2004,RoeBar:2008}, 
especially for the organizations.
While the  projection networks can be useful, the construction of the projections intrinsically 
loses information available in the bipartite networks, which can lead to incorrect community 
structures \citep{GuiSalAma:2007}. In the present work, we thus focus exclusively on representations 
of the Framework Programs
as bipartite networks.

\section{Community Structure} \label{sec:commstruct}

Of great current interest is the identification of community groups,
or modules, within networks.  Stated informally, a community group
is a portion of the network whose members are more tightly linked
to one another than to other members of the network. A variety of
approaches
\citep{AngBocMarPelStr:2007,GolKog:2006,Has:2006,NewLei:2007,ReiBor:2006,PalDerFarVic:2005,NewGir:2004,ClaNewMoo:2004,GirNew:2002}
have been taken to explore this concept; \citet{DanDiaDucAre:2005} 
and \citet{New:2004b} provide useful reviews.
Detecting community groups allows quantitative investigation of
relevant subnetworks. Properties of the subnetworks may differ from
the aggregate properties of the network as a whole, \eg,  modules
in the World Wide Web are sets of topically related web pages.
Thus, identification of community groups within a network is a first
step towards understanding the heterogeneous substructures of the
network.

Methods for identifying community groups can be specialized to
distinct classes of networks, such as bipartite networks
\citep{Bar:2007,GuiSalAma:2007}. This is immediately relevant for our study of the 
FP networks, allowing us to examine the community structure in the bipartite networks. Communities are 
expected to be formed of groups of organizations engaged in \RD into similar topics, and the projects in 
which those organizations take part. 

\subsection{Modularity}

To identify communities, we take as our starting point the
\newterm{modularity}, introduced by \citet{NewGir:2004}. Modularity
makes  intuitive notions of community groups precise by comparing
network edges to those of a null model. The modularity~\( \modularity
\) is proportional to the difference between the number of edges
within communities \( c \) and those for a null model:
\begin{equation}
	\modularity\equiv \frac{1}{2\numedges }\sum_{c}\sum_{i,j\in c}\left(
	\adjelem{i}{j}-\probelem{i}{j}\right) \mathperiod
	\label{eq:modularity}
\end{equation}
Along with \eqn{eq:modularity}, it is necessary to provide a null model, 
defining \( \probelem{i}{j} \). 

The standard choice for the null model 
constrains the degree distribution for the vertices to match the degree 
distribution in the actual network. Random graph models of this sort 
are obtained \citep{ChuLu:2002} by putting an edge between 
vertices \( i \) and \( j \) at random, with
the constraint that on average the degree of any vertex \( i \) 
is \( \vertexdegree{i} \). This constrains the expected adjacency matrix such that 
\begin{equation}
	\vertexdegree{i}=E\left( \sum_{j}\adjelem{i}{j}\right) \mathperiod
\end{equation}
Denote \( E\left( \adjelem{i}{j}\right)  \) by \( \probelem{i}{j} \) and assume further that \( \probelem{i}{j} \)
factorizes into
\begin{equation}
	\probelem{i}{j}=p_{i}p_{j} \mathcomma
\end{equation}
leading to
\begin{equation}
	\probelem{i}{j}\equiv \frac{\vertexdegree{i}\vertexdegree{j}}{2\numedges } \mathperiod
\end{equation}
A consequence of the null model choice is that \( \modularity = 0 \) when all vertices are in 
the same community.

The goal now is to find a division of the vertices into communities such
that the modularity \( \modularity \) is maximal. An exhaustive search for a decomposition
is out of the question: even for moderately large graphs there are far too
many ways to decompose them into communities. Fast approximate algorithms do
exist \citep[see, for example][]{PujBejDel:2006,New:2004a}). 

\subsection{Finding Communities in Bipartite Networks}

Specific classes of networks have additional constraints that can be reflected in the null model.
For bipartite graphs, the null model should be modified to reproduce the
characteristic form of bipartite adjacency matrices:
\begin{equation}
	\adjmat=\bipartsubstructure{\adjsubmat}
	\mathperiod
	\label{eq:bipartsubstructure}
\end{equation}
Recently, specialized modularity measures and search algorithms
have been proposed for finding communities in bipartite networks
\citep{Bar:2007,GuiSalAma:2007}. These measures and methods have
not been studied as extensively as the versions with the standard
null model shown above, but many of the algorithms can be adapted
to the bipartite versions without difficulty. Limitations of
modularity-based methods \citep[\eg, the resolution limit described
by][]{ForBar:2007} are expected to hold as well.

Community identification is then a search for high modularity
partitions of the vertices into disjoint sets. An exhaustive search
for the globally optimal solution is only feasible for the smallest
networks, as the number of possible partitions of the vertices grows
far too rapidly with network size. Several heuristics exist to find
high-quality, if suboptimal, solutions in a reasonable length of
time. For the FP networks, we use a two-stage search procedure:
\begin{enumerate} 
	\item Agglomerative hierarchical clustering, where
	small communities are successively joined into larger ones such
	that the modularity increases. This stage is based on the so-called
	fast modularity (\fm) algorithm \citep{ClaNewMoo:2004}.  
	
	\item Greedy search, where vertices are moved amongst existing communities
	to ensure the resulting partition is at a local optimum of modularity.
	This stage uses the bipartite, recursively induced modules (\brim) algorithm \citep{Bar:2007}.  
\end{enumerate}
The coarse structure is found with \fm, with incremental improvements provided by \brim.

In principle, the above approach should be continued until a maximum in the modularity is 
found. In practice, an excessively large number of communities for
visualization purposes can result, obscuring the core community structure. 
To deal with this difficulty, communities are further merged, so long as the modularity
stays near the maximum (within \( \approx \)90\%) and the general structure of the 
communities is maintained 
\citep[as determined with information theoretic methods, see][]{DanDiaDucAre:2005}. 

\section{Communities in the Framework Program Networks} \label{sec:fpcommunities}

%
%

\begin{figure}
	\includegraphics[width=\textwidth]{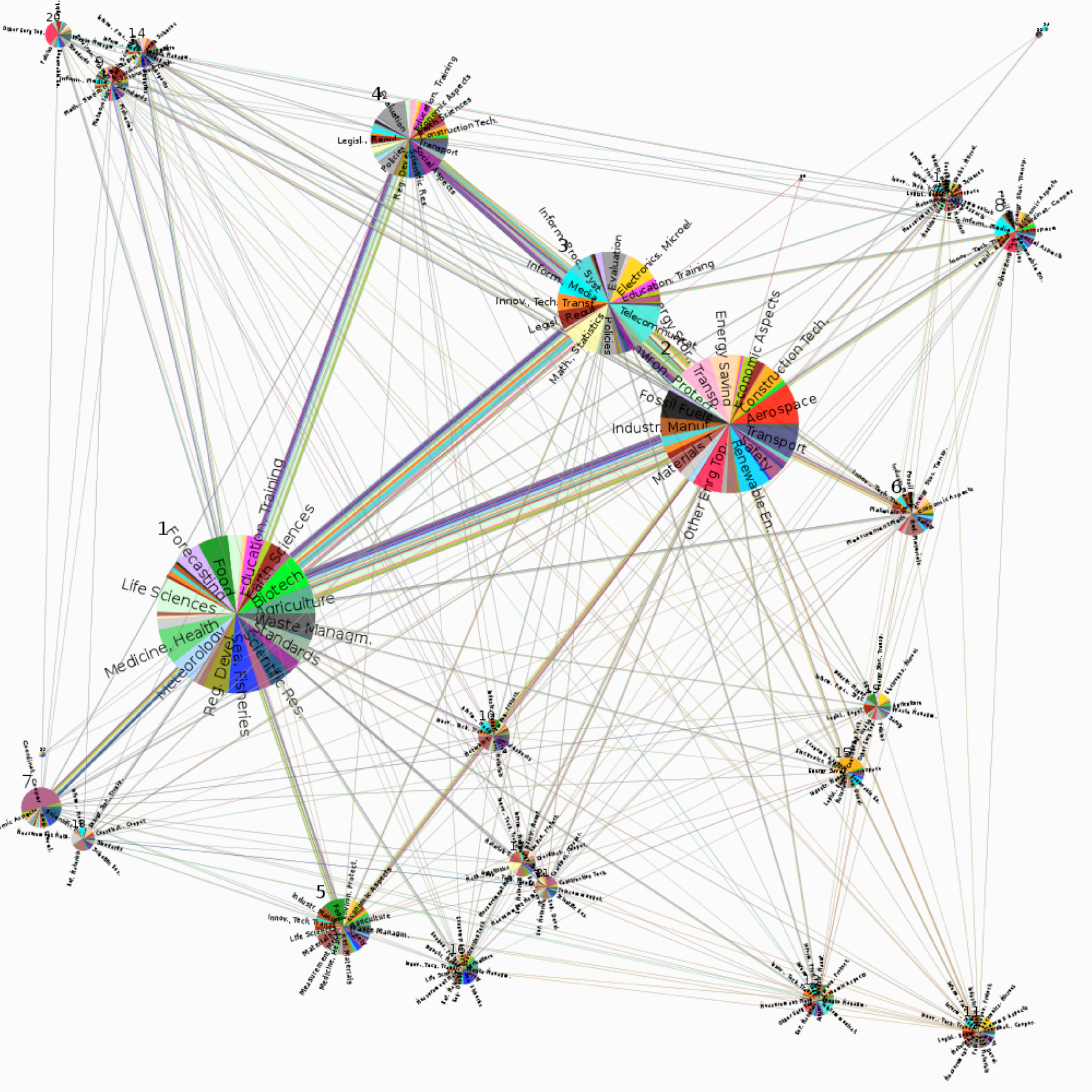}
	\caption{Community groups in the bipartite network of projects and organizations for FP5.}
	\label{fig:brimclusters}
\end{figure}

In \fig{fig:brimclusters}, we show a community structure for FP5,
found as described above, with a modularity of \( \modularity
= 0.644 \) for 25 community groups. The communities are shown as
vertices in a network, with the vertex positions determined using
spectral methods \citep{SeaRic:2003}.  The area of each vertex is
proportional to the number of edges from the original network within
the corresponding community.  The width of each edge in the community
network is proportional to the number of edges in the original
network connecting community members from the two linked groups.
The vertices and edges are shaded to provide additional information
about their topical structure, as described in the next section.
Each vertex is numbered, with numbers
assigned starting from 1   based on the size of the communities, with the largest
communities having the smallest numbers.

The networks from all of the FPs show definite community structure. In each case, the modularity
exceeds 0.6 (see \tbl{tbl:fpcomms}). 

\begin{table}
	\begin{center}
		\begin{tabular}{ccc}
			FP & Number of Communities & Modularity \\ \hline
			2 & 16 & 0.641 \\ 
			3 & 14 & 0.627 \\ 
			4 & 25 & 0.662 \\ 
			5 & 25 & 0.644 \\ 
			6 & 25 & 0.632
		\end{tabular}
		\caption{Communities in the FP networks. In each network, definite community structure is observed.}
		\label{tbl:fpcomms}
	\end{center}
\end{table}

\subsection{Topical Profiles of Communities}

Projects are assigned one or more standardized subject indices. There are 49 subject indices in total, 
ranging from \subjectindex{Aerospace} to \subjectindex{Waste Management}.
We denote by
\begin{equation}
	\freq{t} > 0 
\end{equation}
the frequency of occurrence of the subject index \( t \) in the network, with 
\begin{equation}
	\sum\limits_{t}\freq{t} = 1\mathperiod
\end{equation}
Similarly we consider the projects within one community \( c \) and the frequency 
\begin{equation}
	\topfreq{c}{t}\geq 0 
\end{equation}
of any subject index \(  t \) appearing in the projects only of that community. We
call \( \topfreqnoarg{c} \) the topical profile of community \( c \) to be compared with that of
the network as a whole. 

Topical differentiation of communities can be measured by comparing their
profiles, among each other or with respect to the overall network. This can be done
in a variety of ways \citep{GibSu:2002}, such as by the Kullback ``distance''
\begin{equation}
	D_{c}=\sum\limits_{t} \topfreq{c}{t}\ln \frac{\topfreq{c}{t}}{\freq{t}} \mathperiod
\end{equation}
A true metric is given by
\begin{equation}
	\topicmetric{c} = \sum\limits_{t} \abs{ \topfreq{c}{t}-\freq{t} } \mathcomma
\end{equation}
ranging from zero to two.

\begin{figure}
	\subfigure[Topical profile of community 1. ]{\includegraphics[width=\textwidth]{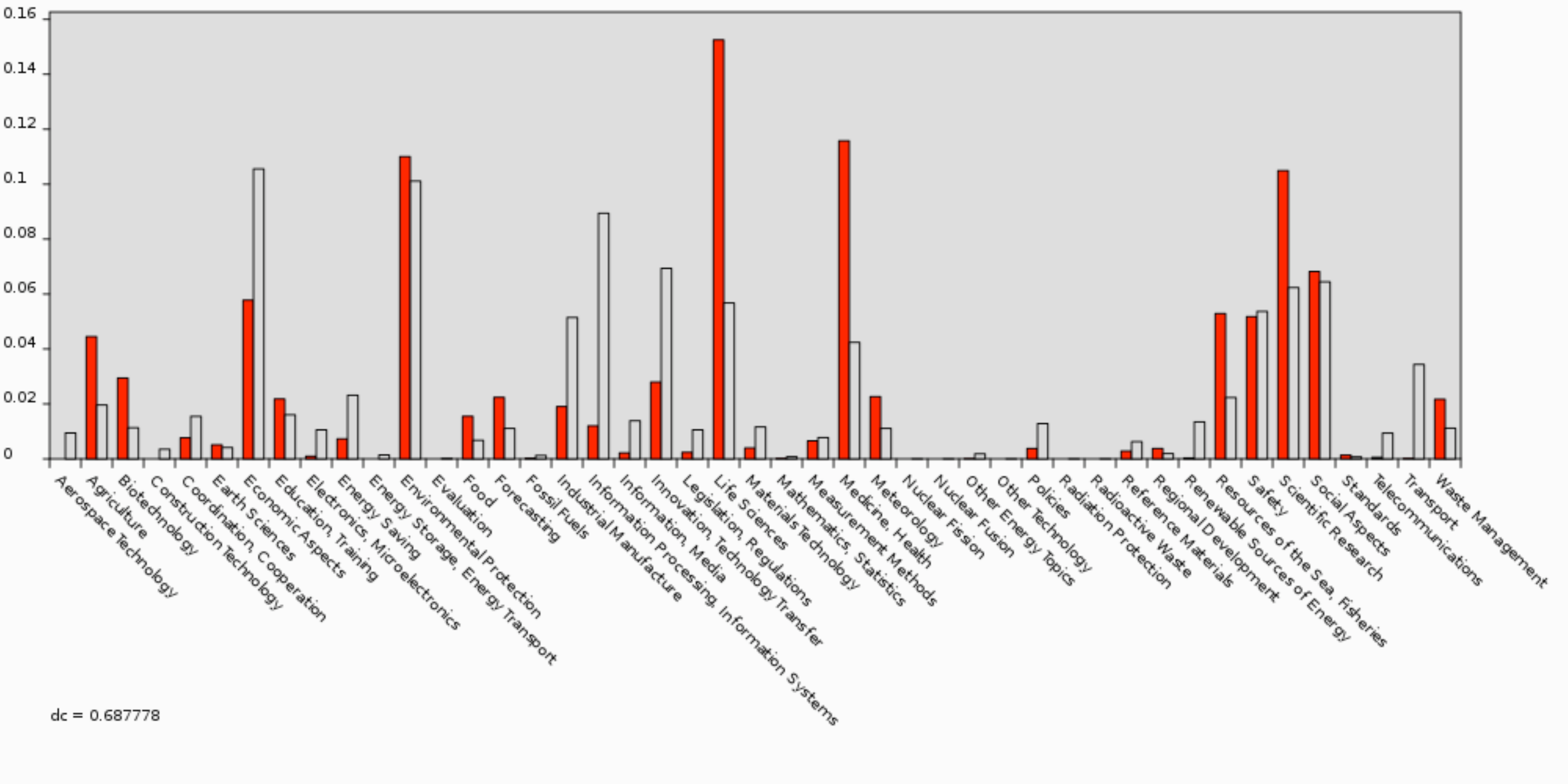} \label{fig:profa}}
	\subfigure[Topical profile of community 2. ]{\includegraphics[width=\textwidth]{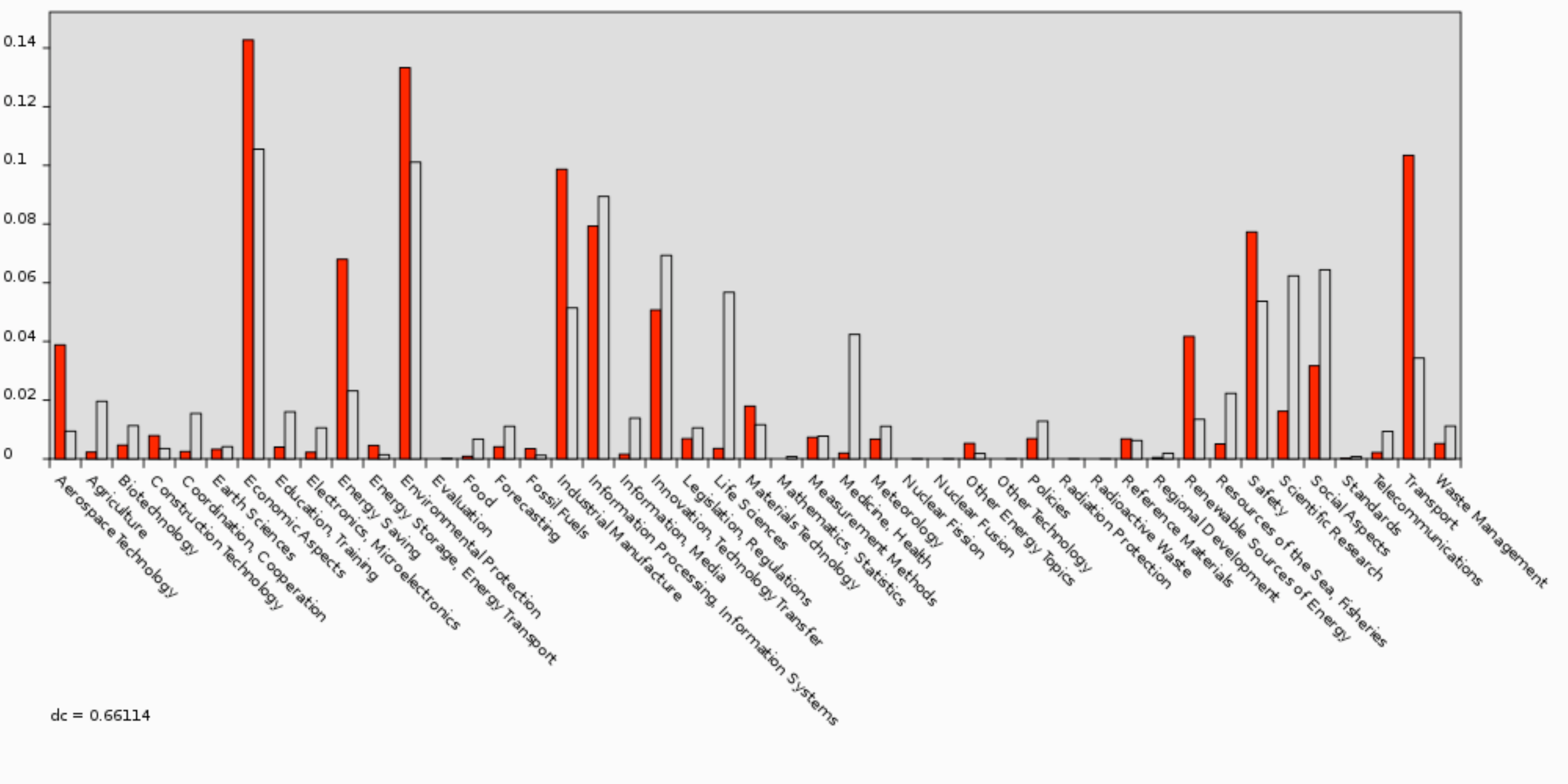}\label{fig:profb} }
	\caption{Community 1 shows strong topical differentiation ($\topicmetric{1} = 0.69$) from the 
	network as a whole, being dominated by topics in  biotechnology and the life sciences. 	
	Community 2 also shows strong topical differentiation ($\topicmetric{2} = 0.66$) from the 
	network as a whole. Further, it is  quite distinct from community 1, being dominated by 
	manufacturing- and transport-related topics.}
\end{figure}

Topical differentiation is illustrated in \figs{fig:profa} \andfig{fig:profb}.
In the figure, example profiles are shown, taken from the network
in \fig{fig:brimclusters}. The community-specific profiles correspond
to the communities 1 and 2  in \fig{fig:brimclusters}.
Both communities are topically differentiated from the network as a whole, and with 
similar extent (\( \topicmetric{1} = 0.69 \), \( \topicmetric{1} = 0.66 \)). However, the 
actual topics are quite different, with community 1 dominated by subject indices relating to
biotechnology and the life sciences, while community 2 is dominated by subject indices 
relating to manufacturing and transport.

\section{Discussion} \label{sec:discussion}

We have presented an investigation of networks derived from the
European Union's Framework Programs for Research and Technological
Development. The networks are of substantial size, complexity, and 
economic importance. We have attempted to provide a coherent picture of the
complete process, beginning with data preparation and network
definition, then continuing with analysis of the network community structure. 

We first considered the challenges involved in dealing
with a large amount of imperfect data, detailing the tradeoffs made
to clean the raw data into a usable form under finite resource
constraints.  The processed data was  used to define bipartite
networks with vertices consisting of all the projects and organizational subentities
involved in each FP. 

Next we analyzed the community structure of the Framework Programs. 
Using a modularity measure and search algorithm adapted to bipartite networks, we 
identified communities from the networks. We found that the communities are topically
differentiated based on the standardized subject indices for Framework Program projects. 

The communities identified will serve as basis for further studies of the Framework 
Programs. A natural extension of the present work is to examine other properties by 
which the communities are differentiated. Properties of organizations making up the 
communities can be explored, much as the subject indices for the projects were 
examined in this work. Immediate candidates for consideration include the types of the 
organizations (\eg, universities or firms) and geographical location of the organizations (\eg,
as countries or using NUTS classifications). Further, the communities will be used as a basis
for modeling determinants of partner choice, such as with spatial interaction models 
\citep{SchBar:2008a,SchBar:2008} or binary choice models \citep{PaiSch:2008}, providing
insight into the formation rules at work in heterogeneous subsets of the Framework Programs.

\acknowledgments

The authors gratefully acknowledge financial support from the European FP6-NEST-Adventure Programme, under contract number~028875, and from the Portuguese FCT, under projects POCTI/MAT/58321/2004/FSE-FEDER and FCT/POCTI-219/FEDER.


\bibliographystyle{plainnat}

\bibliography{references}

\end{document}